\documentclass[twocolumn,showpacs,amsfonts,aps,prc,floatfix,superscriptaddress,nofootinbib]{revtex4}

\usepackage{amsmath}
\usepackage{bm}
\usepackage{graphicx}

\newcommand{\ecc}{\varepsilon}

\newcommand{\be}[1]{\begin{equation}\label{#1}}
\newcommand{\ee}{\end{equation}}

\newcommand{\eq}{{\,=\,}}

\usepackage{color}

\begin{document}

%%%%%%%%%%%%%%%%%%%%%%%%Front Matter%%%%%%%%%%%%%%%%%%%%%%%%%%%%%%%%%%
%%%%%%%%%%%%%%%%%%%%%%%%%%%%%%%%%%%%%%%%%%%%%%%%%%%%%%%%%%%%%%%%%%%%%%

\title{Elliptic flow in $\sqrt{s}=200\ \mathrm{GeV}$ Au+Au collisions
and $\bm{\sqrt{s}=2.76}$\,TeV Pb+Pb collisions: insights from viscous 
hydrodynamics + hadron cascade hybrid model}

\author{Huichao Song}
\email[Correspond to\ ]{HSong@lbl.gov}
\affiliation{Nuclear Science Division,
             %MS 70R0319,
             Lawrence Berkeley National Laboratory, Berkeley,
             California 94720, USA}

\author{Steffen A. Bass}
\affiliation{Department of Physics, Duke University, Durham,
             North Carolina 27708, USA}

\author{Ulrich Heinz}
\affiliation{Department of Physics, The Ohio State University,
             %191 West Woodruff Avenue,
             Columbus, Ohio 43210, USA}

\begin{abstract}
Using the newly developed hybrid model {\tt VISHNU} which connects viscous 
hydrodynamics with a hadron cascade model, we study the differential and 
integrated elliptic flow $v_2$ at different centrality bins for 200 A GeV 
Au+Au collisions and 2.76 A TeV Pb+Pb collisions. We find that the 
\emph{average} Quark Gluon Plasma (QGP) specific shear viscosity $\eta/s$ 
slightly increases from Relativistic Heavy Ion Collider (RHIC) to Large
 Hadron Collider (LHC) energies. However, a further study assuming different 
temperature dependencies for $(\eta/s)_\mathrm{QGP}$ shows that one cannot 
uniquely constrain the form of $(\eta/s)_\mathrm{QGP}(T)$ by fitting the 
spectra and $v_2$ alone. Based on our current understanding, the question 
whether the QGP fluid is more viscous or more perfect in the temperature 
regime reached by LHC energies is still open.
\end{abstract}

\date{\today}

\maketitle

%%%%%%%%%%%% begin text %%%%%%%%%%%%%%%%%%%
The first heavy-ion data from the Large Hadron Collider (LHC) have revealed 
many phenomena that are very similar to those seen at the Relativistic 
Heavy-Ion Collider (RHIC) at lower beam energies, such as elliptic flow and 
jet energy-loss. The question arises whether the increased reach in 
temperature and energy-density attainable at the LHC allows for the 
identification of systematic trends, for example in the temperature 
dependence of the specific shear viscosity $\eta/s$ of the produced 
Quark-Gluon Plasma (QGP).

Measurements by the ALICE collaboration for $2.76\,A$\,TeV Pb+Pb collisions 
shows that the charged hadron multiplicity density is about a factor of 
2.2 higher than the one from 200 A GeV Au+Au collisions
\cite{Aamodt:2010pb,Aamodt:2010cz}. Assuming a similar QGP thermalization 
time at RHIC and LHC energies and a linear relationship between the final 
multiplicity and initial entropy, one finds that the initial temperature 
of the QGP fireball at lower LHC energies is about 30\% larger than the 
one at top RHIC energies. Meanwhile, the differential elliptic flow 
$v_2(p_T)$ of charged hadrons measured by the ALICE collaboration as a 
function of transverse momentum $p_T$ (using the 4-particle cumulant method) 
is, up to transverse momenta of 3 GeV/$c$, nearly identical to that 
measured by the STAR collaboration at RHIC, independent of collision 
centrality \cite{Aamodt:2010pa}. When integrated over the transverse 
momentum, on the other hand, the total $v_2$ from ALICE is about 
30\% higher than that from STAR, due to an increase of the mean $p_T$ 
for the LHC spectra \cite{Aamodt:2010pa}.

The larger integrated charged hadron elliptic flow at the LHC implies 
a larger total momentum anisotropy \cite{Song:2010mg,Song:2011hk} and 
suggests a higher efficiency of the QGP fluid for converting the initial 
spatial deformation of the collision fireball into anisotropic collective 
flow. This raises the following questions: is the QGP fluid still strongly 
coupled at LHC energies? Does the QGP fluid have a similar specific shear 
viscosity at RHIC and LHC energies? In this article, we shall address the 
latter question.

Several groups \cite{Luzum:2010ag,Bozek:2011wa,Schenke:2011tv} have 
recently published analyses of the ALICE data \cite{Aamodt:2010pa} based 
on a purely hydrodynamic approach, and another group offered an assessment 
of the same data within an ideal hydrodynamic + hadron cascade hybrid 
approach \cite{Hirano:2010je}. In contrast, we here use the newly developed 
hybrid approach {\tt VISHNU} \cite{Song:2010aq} which correctly describes
viscous and other dissipative effects in both the early QGP and late
hadronic rescattering stages, including the breakdown of chemical 
equilibrium in the hadron gas \cite{Hirano:2010je}, by connecting the 
(2+1)-dimensional {\em viscous} hydrodynamic code {\tt VISH2+1} 
\cite{Song:2007fn} with the microscopic hadronic transport model {\tt UrQMD} 
\cite{Bass:1998ca} (see Refs.~\cite{Song:2010aq,Song:2007fn,Bass:1998ca} 
for details). Comparing to a pure hydrodynamic 
approach \cite{Luzum:2010ag,Bozek:2011wa,Schenke:2011tv}, the kinetic and 
chemical freeze-out of the system is naturally described by the microscopic 
hadronic evolution part of VISHNU through 
elastic, semi-elastic and inelastic scatterings, which eliminates the 
corresponding freeze-out parameters used in purely hydrodynamic calculations. 
The transition from the viscous hydrodynamic to the microscopic evolution 
parts of the model occurs at the switching temperature, $T_\mathrm{sw}$, 
which is chosen to be 165\,MeV, i.e.  near the QGP phase transition 
temperature $T_\mathrm{c}$ and adjusted to reproduce the chemical 
freeze-out temperature at RHIC energies~\cite{BraunMunzinger:2001ip}. 
This is the highest $T_\mathrm{sw}$ possible for a hadronic transport 
description as well as the lowest $T_\mathrm{sw}$ possible for 
hydrodynamics without introducing additional parameters related to 
hadronic non-chemical equilibrium and dissipative effects in the 
hydrodynamic part~\cite{Song:2010aq}. For the hydrodynamic evolution 
above $T_\mathrm{sw}$, we use the equation of state (EOS) {\tt s95p-PCE}
\cite{Shen:2010uy} which is based on recent lattice results 
\cite{Bazavov:2009zn}.

%
%============================ Fig. 1 ==================================
\begin{figure*}[t]
\includegraphics[height=6.2cm,clip=,angle=0]{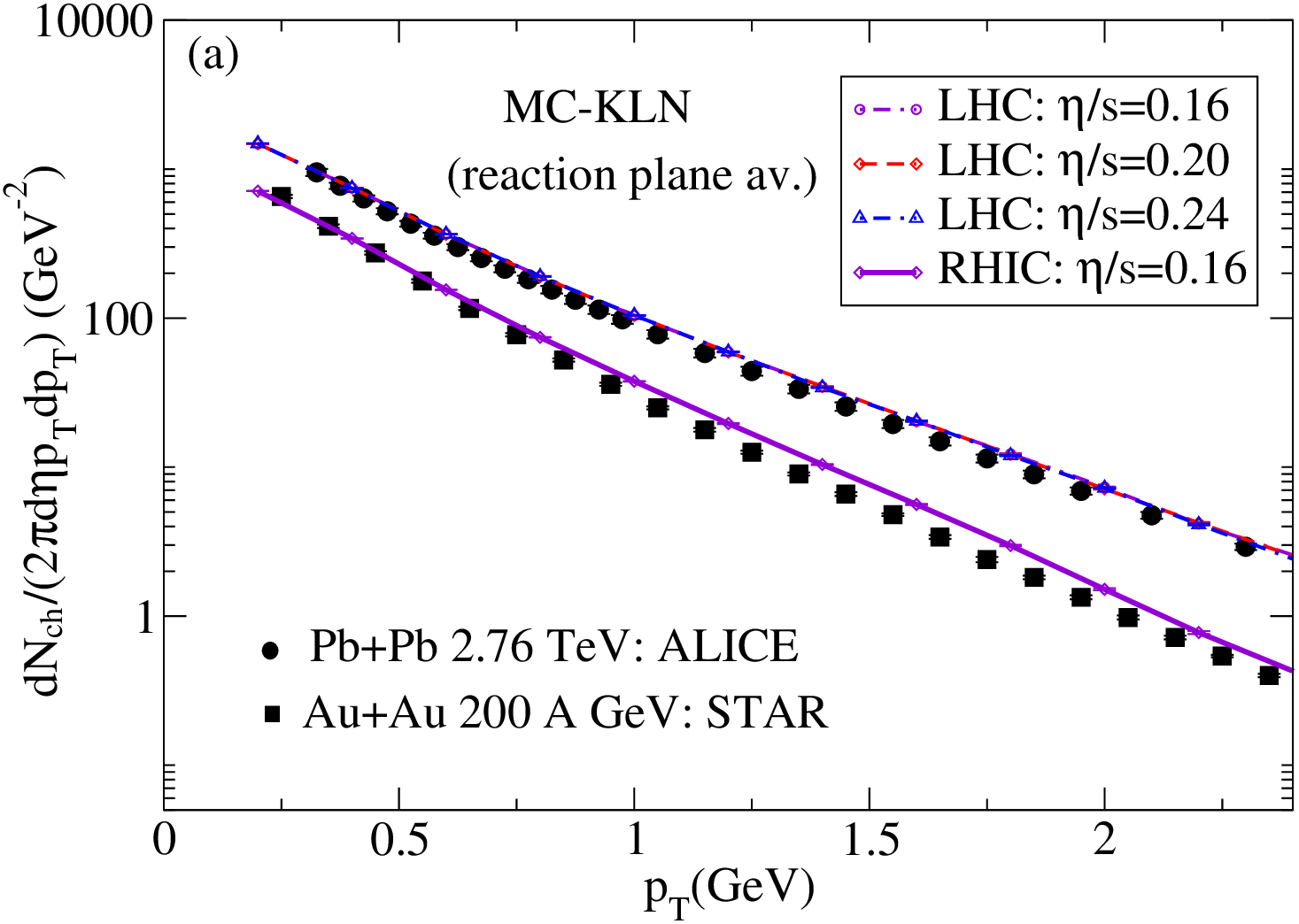}
\includegraphics[height=6.1cm,clip=,angle=0]{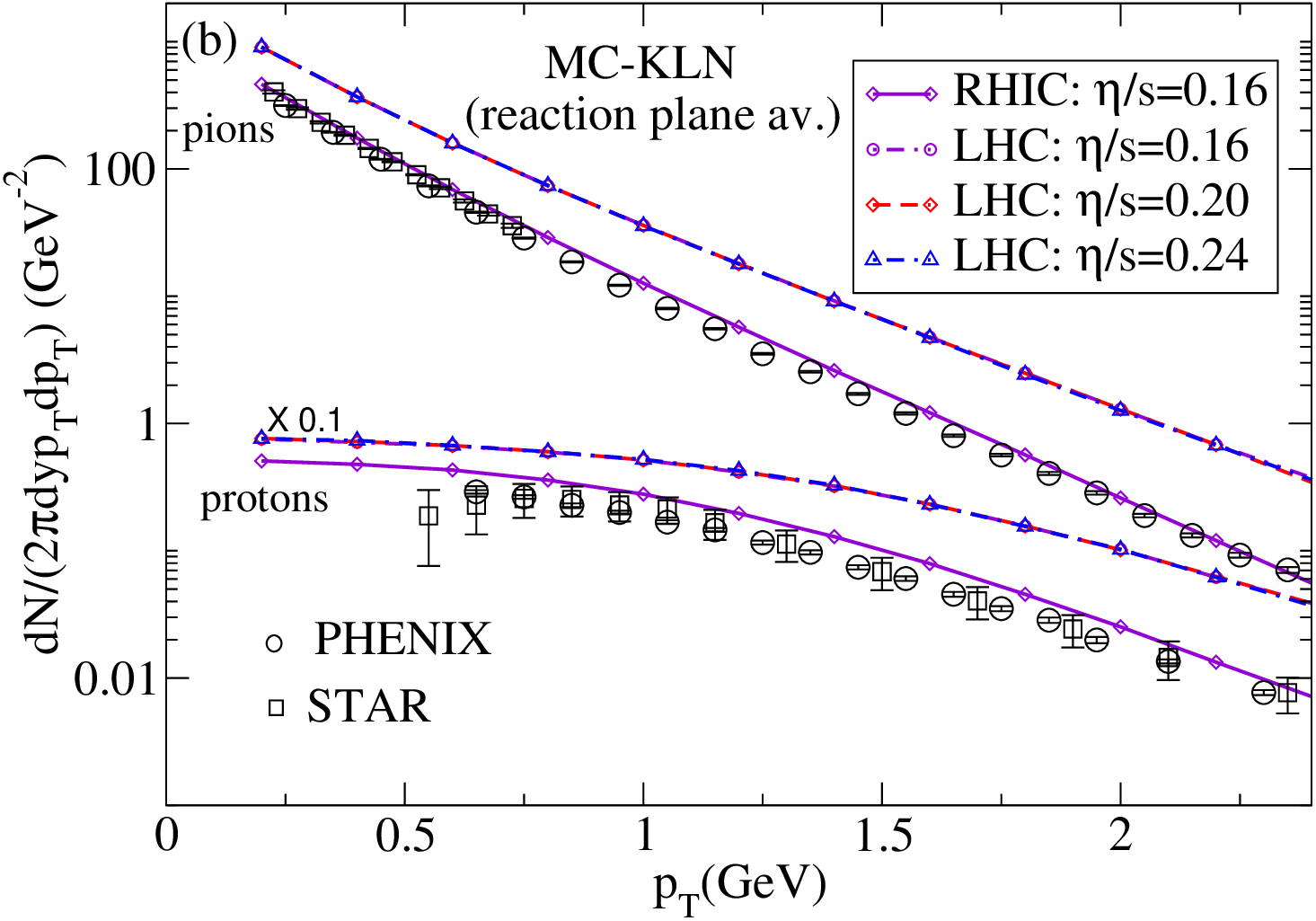}
\caption{\label{F1} (Color online) Left: $p_T$ spectra for all charged 
hadrons. Experimental data are from ALICE~\cite{Aamodt:2010jd} and STAR
\cite{Adams:2003kv}. Right: $p_T$ spectra for pions and protons. 
Experimental data are from STAR~\cite{Adams:2003xp} and PHENIX
\cite{Adler:2003cb}. }
\end{figure*}
%======================================================================
%

In this manuscript, we will focus on the specific shear viscosity of the QGP
at RHIC and LHC energies, neglecting the bulk
viscosity\footnote{Ref.~\cite{Song:2009rh} shows that the bulk viscosity
   results in less than 20\% contamination for the extracted
   QGP shear viscosity value, due to the critical slowing down of the bulk
   relaxation time near $T_\mathrm{c}$.}
and assuming zero net baryon density and heat conductivity.
In our calculations we shall either use a constant $\eta/s$ at RHIC and LHC 
energies (which represents the \emph{averaged} shear viscosity effects 
during the whole evolution of the QGP fireball) or input different 
temperature dependencies for $(\eta/s)_\mathrm{QGP}(T)$ in the temperature 
range covered by both RHIC and LHC energies. The corresponding relaxation 
time is set as $\tau_\pi = 3 \eta / ( s T)$.\footnote{The choice of 
  $\tau_\pi$ in the QGP phase has small to negligible effects on the 
  final spectra and elliptic flow~\cite{Song:2007fn}.}

In order to start a {\tt VISHNU} calculation, initial conditions are 
required. The two most popular geometric models for initial particle 
production including fluctuation effects for high-energy heavy ion 
collisions are the Monte-Carlo Glauber Model (MC-Glauber) 
\cite{Miller:2007ri,Hirano:2009ah} and the Monte-Carlo KLN Model (MC-KLN)
\cite{Miller:2007ri,Drescher:2006ca}. Following 
Ref.~\cite{Hirano:2009ah,Song:2010mg,Song:2011hk}, we account for 
event-by-event fluctuations on average by using an initial entropy 
density profile that is the result of averaging over a large number of 
fluctuating initial entropy density distributions. Such an average can 
be done in two ways: either by re-centering and rotating each Monte-Carlo 
event to align the major and minor axes of each initial density distribution 
before averaging (\emph{initialization in the participant plane}) or by 
directly averaging without re-centering and rotating (\emph{initialization 
in the reaction plane}). In this paper, we use the reaction plane method 
since we will compare our theoretical results with the elliptic flow data 
extracted by STAR and ALICE using the 4-particle cumulant method, 
$v_2\{4\}$, which minimizes non-flow effects \cite{Ollitrault:2009ie} 
and measures $v_2$ in the reaction plane under the assumption of Gaussian 
fluctuations \cite{Bhalerao:2006tp}. The assumption of Gaussian 
fluctuations was challenged in Ref.~\cite{Alver:2008zza} for the most 
central and peripheral collisions; a recent analysis \cite{Qiu:2011iv}
confirmed, however, that for collision centralities up to about 40\%
reaction-plane averaged initial conditions produce initial eccentricities
that are very close to $\ecc\{4\}$, the presumed driver for $v_2\{4\}$ 
\cite{Bhalerao:2006tp}. Only for more peripheral collisions does this
shortcut break down \cite{Qiu:2011iv}.

Questions about the validity of the direct comparison of the 
reaction-plane averaged theoretical results with $v_2\{4\}$ data from 
peripheral collisions engender corresponding uncertainties for the 
extracted values of the QGP viscosity. Additional $5{-}10\%$ uncertainties 
arise from the recent observation \cite{Andrade:2008xh,Schenke:2010rr,%
Qiu:2011iv} (see also \cite{Petersen:2010md}) that single-shot hydrodynamic 
evolution of a smooth averaged initial profile (as employed here) slightly 
overpredicts the elliptic flow compared to an event-by-event hydrodynamic 
evolution of each fluctuating and highly inhomogeneous initial profile
separately where the average over the event ensemble is taken only at the 
end. In contrast to Ref.~\cite{Song:2010mg,Song:2011hk}, we here do not 
aim to extract the QGP viscosity at RHIC and LHC energies with reliable 
uncertainty estimates, but rather to investigate the relative change of 
the QGP viscosity from RHIC to LHC energies. Assuming that 
distortions from non-Gaussian eccentricity fluctuations and the replacement
of event-by-event evolution by single-shot hydrodynamics are similar for the 
STAR and ALICE data, their presence will not affect the conclusions regarding 
the variation of the QGP viscosity with collision energy drawn from a 
comparison of $v_2\{4\}$ data with our calculations. For the same reason 
and for the sake of computing efficiency, we only utilize the MC-KLN 
initialization in the reaction plane,\footnote{Studies show that the 
choice of MC-Glauber or MC-KLN initializations will only affect the 
absolute value of the extracted specific shear viscosity at RHIC and 
LHC energies, but that the trend on how the QGP viscosity changes from 
RHIC to LHC is very similar for MC-Glauber and MC-KLN initializations
\cite{Song-Prv}.} and leave the case for a detailed extraction of QGP 
viscosity at LHC energies to future studies.

%
%============================ Fig. 2 ==================================
\begin{figure}[h]
\includegraphics[height=6.2cm,clip=,angle=0]{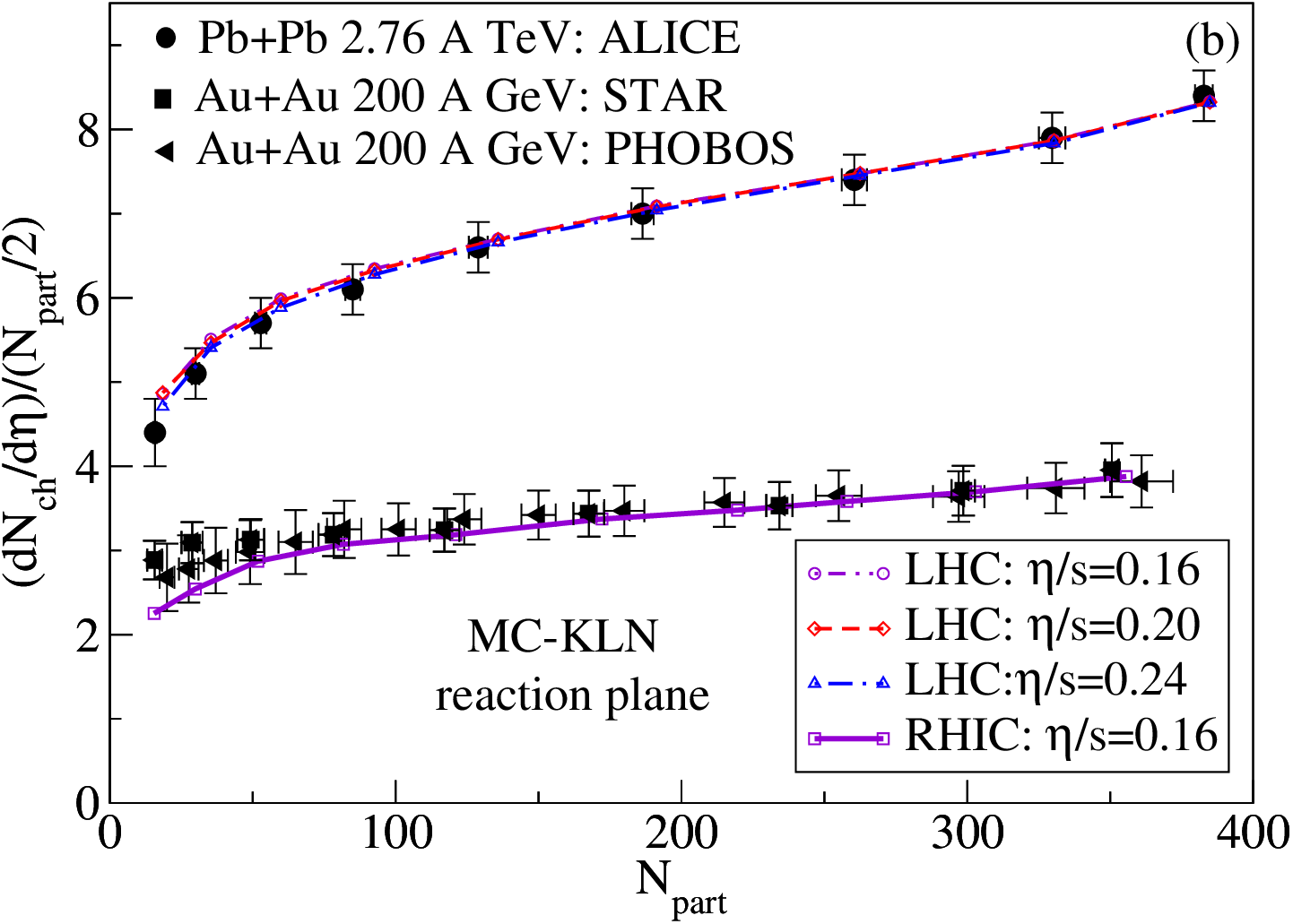}
\caption{\label{F1b} (Color online) Centrality dependence of the charged 
hadron pseudo-rapidity density per participant pair 
$(dN/d\eta)/(N_\mathrm{part}/2)$. Experimental data are from ALICE
\cite{Aamodt:2010cz}, STAR \cite{:2008ez} and PHOBOS \cite{Back:2004dy}.
Theoretical lines in both panels are from {\tt VISHNU} with different 
constant $\eta/s$ as input (see the text for the details of other inputs 
and parameters).}
\end{figure}
%======================================================================
%

The initial time $\tau_0$ and the normalization of the averaged initial 
entropy density profile need to be fixed from experimental data. Following 
Refs.~\cite{Song:2010mg,Song:2011hk}, we use the following parameter sets 
for the shear viscosity to entropy ratio $\eta /s$ and hydrodynamic 
starting time $\tau_0$: (0.16, 0.9 fm/c), (0.20, 1.05 fm/c) and 
(0.24, 1.2 fm/c). Please note that for a larger value of the QGP 
viscosity, we use a later starting time $\tau_0$ to compensate for the 
additional radial flow generated by that larger viscosity
\cite{Song:2010mg,Romatschke:2007jx}. After tuning the normalization of 
the initial entropy density to approximately reproduce the final state 
charged hadron multiplicity per unit of pseudo-rapidity in 200 A GeV 
central Au+Au collisions ($dN/d\eta \simeq 690$~\cite{:2008ez,Back:2004dy}) 
and in 2.76 A TeV central Pb+Pb collisions ($dN/d\eta \simeq 1600$
\cite{Aamodt:2010pb}), we find that our calculation provides a good 
description of the data on $p_T$ spectra for all charged hadrons in most 
central collisions for STAR~\cite{Adams:2003kv,:2008ez} and ALICE
\cite{Aamodt:2010jd} as shown in Fig.~\ref{F1}a.  Ref.~\cite{Song:2011hk} 
also shows that with the above parameters one can obtain a good fit to 
the $p_T$-spectra for identified hadrons (such as pions and protons) 
from most central collision to most peripheral collisions at RHIC energies. 
We find that, with the above adjustment of the starting time $\tau_0$ 
when changing $\eta/s$, these $p_T$-spectra are rather insensitive to 
the QGP viscosity. In Fig.~\ref{F1}b, we show the $p_T$-spectra for 
pions and protons in most central collisions, and compare the RHIC results 
with the STAR \cite{Adams:2003xp} and PHENIX data \cite{Adler:2003cb}. 
Due to the current lack of ALICE data for identified hadrons, the 
corresponding LHC results are predictions.

%
%============================ Fig. 3 ==================================
\begin{figure}[b]
\includegraphics[width=0.95\linewidth,clip=,angle=0]{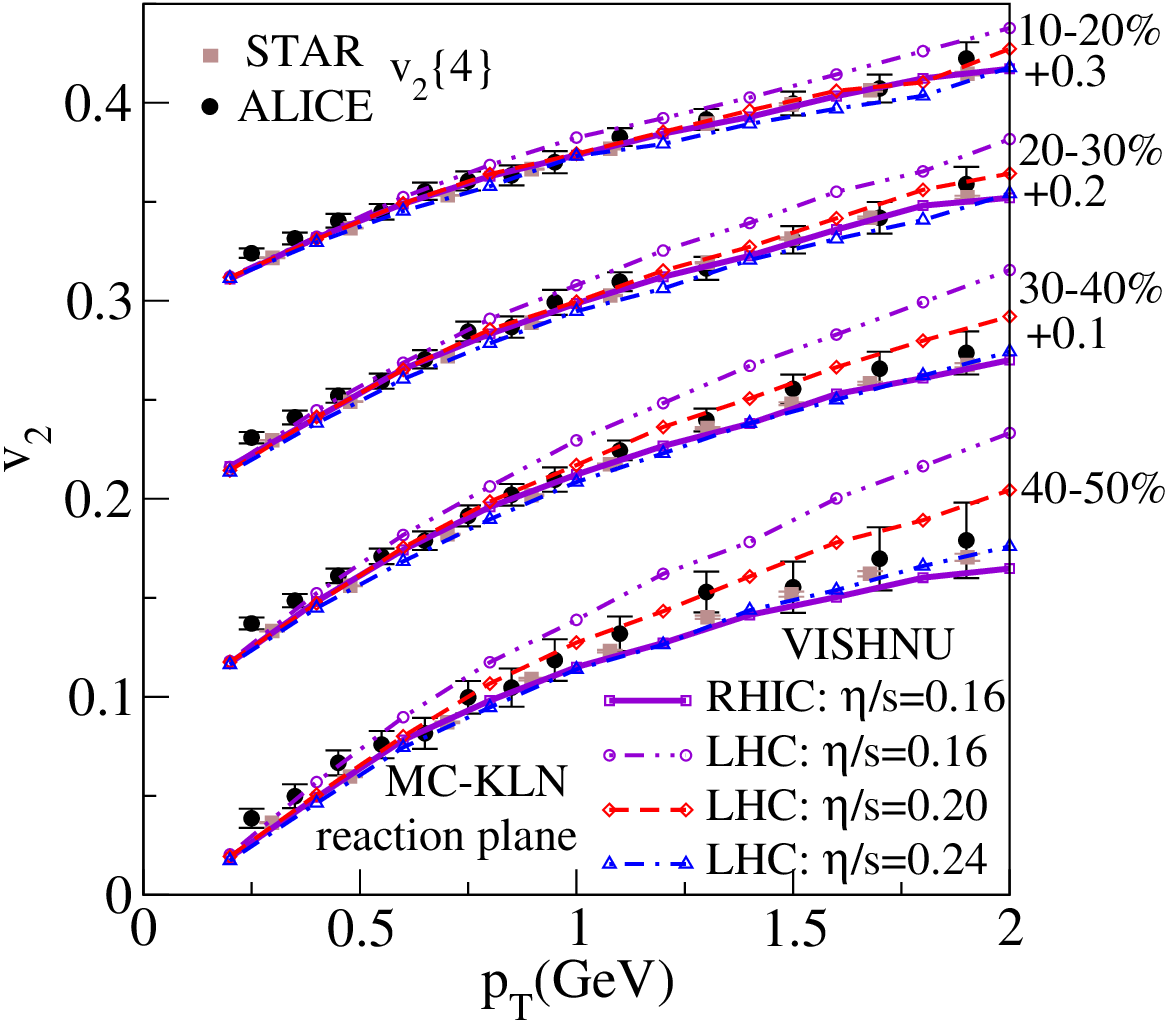}
\caption{\label{F2} (Color online) $v_2(p_T)$ at 10-20\%, 20-30\%, 30-40\% 
and 40-50\% centrality. Experimental data are from STAR~\cite{:2008ed} and 
ALICE~\cite{Aamodt:2010pa} obtained from 4 particle cumulant method. 
Theoretical lines are from {\tt VISHNU} calculations with different 
constant $\eta/s$ as input. See text for details.}
\end{figure}
%======================================================================
%

In the MC-KLN initialization, we use the standard parametrization for the 
saturation scale $Q^2_{s,A}$ as shown in~\cite{Hirano:2009ah}, which is 
tuned to reproduce the centrality dependence of the charged hadron 
multiplicity for 200 A GeV Au+Au collisions. Fig.~\ref{F1b} shows that 
such  aparametrization also leads to a good description for the slope of 
the $(dN_\mathrm{ch}/d\eta)/(N_{part}/2)-N_{part}$ curve in 2.76 A TeV 
Pb+Pb collisions. However, we have to point out that the same 
parametrization of $Q^2_{s,A}$ will lead to a slight overprediction
of the value of $dN_\mathrm{ch}/d\eta$ at LHC energies as shown in 
Ref. \cite{Aamodt:2010cz}. To avoid the over-generation of elliptic flow 
from over-predicted final multiplicities, we tune the normalization of 
initial entropy density as described above to fit the $dN/d\eta$ in the 
$0{-}5\%$ centrality bin. This leads to a good fit on the overall magnitude 
of the $(dN_\mathrm{ch}/d\eta)/(N_\mathrm{part}/2)$ vs. $N_\mathrm{part}$ 
curve.

Having fixed all parameters, we calculate the differential elliptic flow
at RHIC and LHC energies for different constant values of $\eta/s$ as input. 
Fig.~\ref{F2} shows that with $\eta/s=0.16$, {\tt VISHNU} nicely fits the 
STAR $v_2 (p_T)\{4\}$ data from 0 to 2 GeV for different centrality bins. 
In contrast, the same  $\eta/s=0.16$ significantly over-shoots the ALICE 
$v_2\{4\}$ data (for $p_T > 0.5$\,GeV/$c$). After increasing $\eta/s$ to 
$0.20 - 0.24$, {\tt VISHNU} can roughly fit the ALICE data at higher $p_T$, 
but still under-predicts the data for  $p_T < 0.5$\,GeV/$c$. This effect 
of under-prediction of the low $p_T$ data is also found in other 
hydrodynamics-based calculations, including the (3+1)-d ideal hydrodynamics + 
hadron cascade simulations by Hirano et. al~\cite{Hirano:2010je}, the (2+1)-d 
viscous hydrodynamic calculations by Bozek~\cite{Bozek:2011wa} and the 
event-by-event simulations with a (3+1)-d viscous hydrodynamics with 
fluctuating initial conditions done by Schenke {\it et al.} 
\cite{Schenke:2011tv}. In Ref.~\cite{Bozek:2011wa},  the deviation is 
interpreted to be due to non-thermalized particles stemming from jet 
fragmentation. While the origin of this deviation  is still under debate, 
we conclude from Fig.~\ref{F2} that one needs a larger averaged QGP 
specific viscosity to fit the ALICE $v_2(p_T)$ at $p_T>0.5$\,GeV/$c$ 
than the one used to fit the corresponding STAR data. This conclusion 
rests on the assumption that the theoretical model correctly describes 
the slopes of the charged hadron $p_T$-spectra at all the centralities 
shown in Fig.~\ref{F2}.

Using {\tt VISHNU} we find that, even at LHC energies where almost all 
of the final momentum anisotropy is generated hydrodynamically in the 
QGP stage, the charged hadron $v_2(p_T)$ continues to grow somewhat 
during the hadronic stage. This hadronic increase of $v_2(p_T)$ is
smaller at the LHC than at RHIC, in agreement with earlier findings 
\cite{Hirano:2005xf} using an ideal hydrodynamic + cascade hybrid code.
At RHIC energies, some of this hadronic increase is driven by the
creation of additional overall momentum anisotropy which has not yet 
quite saturated in the QGP phase. At LHC energies, it is mostly caused
by a hadronic redistribution of the momentum anisotropy already established 
in the QGP phase in $p_T$ and among the various different hadronic species,
due the hadronic increase in radial flow that pushes $v_2$ to larger
$p_T$, especially for heavy particles \cite{Huovinen:2001cy}. This
effect is sensitive to the chemical composition in the hadron gas
\cite{Hirano:2002ds,Kolb:2002ve,Hirano:2005wx,Huovinen:2007xh}, and
a corresponding hadronic increase of $v_2(p_T)$ is not observed in purely 
hydrodynamic calculations with an equation of state that (incorrectly) 
assumes chemical equilibrium among the hadrons even below the chemical 
decoupling temperature $T_\mathrm{chem}{\,\approx\,}165$\,MeV 
\cite{Schenke:2011tv}.

\begin{figure}[t]
\includegraphics[width=0.99\linewidth,clip=,angle=0]{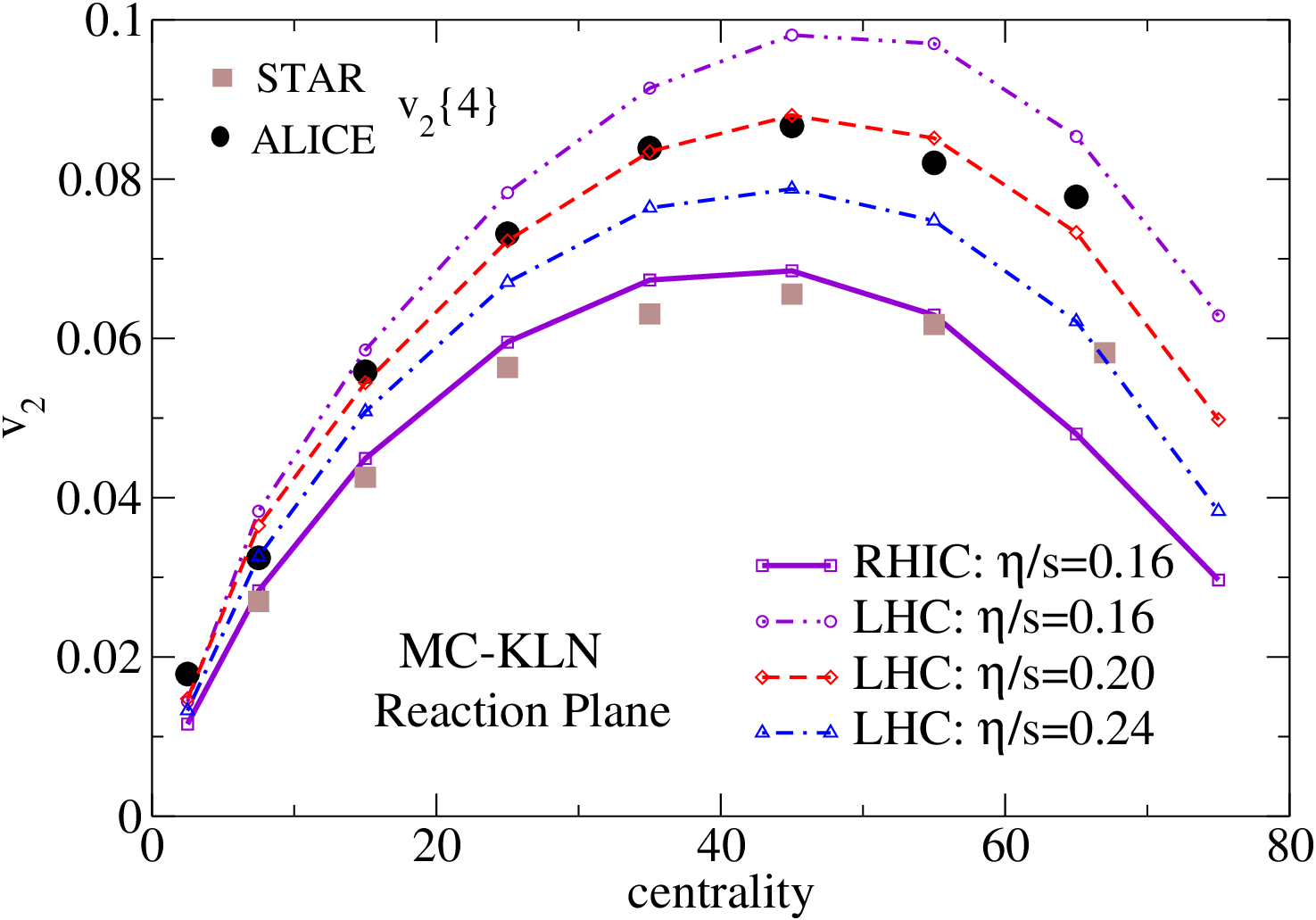}
\caption{\label{F3}(Color online) Integrated $v_2$ as a function of
centrality. Same illustrations for theoretical and experiential lines as 
in Fig.~\ref{F2}.}
\end{figure}

Fig.~\ref{F3} shows the comparison of experimental and theoretical 
integrated $v_2$, obtained from integrating $v_2 (p_T)$ with the 
corresponding $p_T$ spectra as weighting functions. Following the 
STAR~\cite{:2008ed} and ALICE~\cite{Aamodt:2010pa} analysis, we use the 
same $p_T$ and pseudo-rapidity cut in our {\tt VISHNU} calculations 
($0.15<p_T<2$\,GeV/$c$ and $|\eta|<1$ at RHIC energy, and 
$0.2<p_T<5$\,GeV/$c$ and $|\eta|<0.8$ at LHC energy). One finds that 
{\tt VISHNU} is capable of fitting the experimental data with $\eta/s=0.16$ 
at RHIC  and $\eta/s=0.20$ at LHC, except for the most peripheral centrality 
bins. Comparing our calculation to the  STAR $v_2(p_T)$ data with 
$\eta/s=0.16$, the solid purple curve with square symbols is slightly 
above the STAR data due to the slight overprediction of the $p_T$-spectra 
around 1 GeV as shown in Fig.~\ref{F1}a. Similarly, the value of 
$\eta/s=0.20$ from the fit to the ALICE integrated $v_2$ is slightly 
below the extracted value of $\eta/s=0.22$ from ALICE $v_2(p_T)$, mainly 
because of the under-fitting of the ALICE $v_2(p_T)$ at lower $p_T$.

In Ref.~\cite{Song:2011hk}, we discussed that integrated $v_2$ is better 
suited than differential $v_2$ for the extraction of the QGP viscosity,  
due to it being directly related to the fluid momentum anisotropy and 
insensitive to other details of hydrodynamic calculation such as chemical 
components of the hadronic phase, the form of non-equilibrium distribution 
function $\delta f$, bulk viscosity and so on. However, due to the current 
deviation between calculations and the ALICE $v_2(p_T)$ data at lower $p_T$, 
which translates into corresponding errors for the integrated $v_2$, any 
extraction of the QGP viscosity from the integrated $v_2$ measurements 
alone at LHC energies cannot be considered robust.

%
%============================ Fig. 4 ==================================
%======================================================================
%
Although both Figs.~\ref{F2} and \ref{F3} indicate that the \emph{averaged} 
QGP specific viscosity (constant $\eta/s$) slightly increases with collision 
energies, it has to be pointed out that using one constant value of $\eta/s$ 
to fit RHIC data and a different constant value of $\eta/s$ to fit LHC data 
is not logically consistent. In other words, one can not describe the QGP 
fluid created at at RHIC energies with $\eta/s\eq0.16$ 
($T_\mathrm{c}{\,<\,}T{\,<\,}2T_\mathrm{c}$) and then use $\eta/s\eq0.22$ 
($T_\mathrm{c}{\,<\,}T{\,<\,}3T_\mathrm{c}$) for the one created at LHC 
energies.  It is a temperature dependent $\eta/s (T)$ that reflects the 
intrinsic properties of the QGP fluid, and this temperature dependence
should be unique and describe the data both at RHIC and LHC energies.

%============================ Fig. 5 ==================================
\begin{figure}[b]
\includegraphics[width=0.9\linewidth,clip=,angle=0]{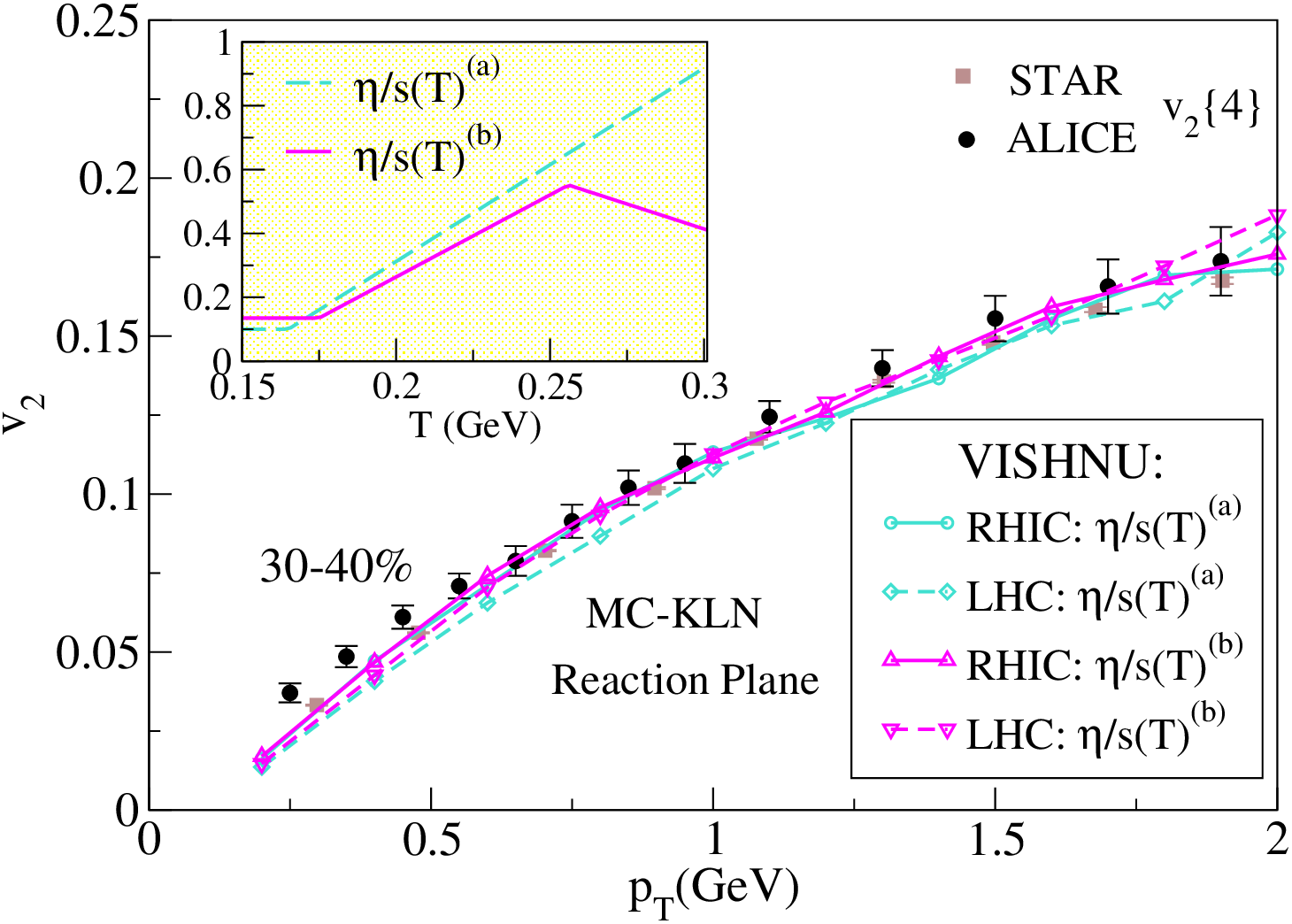}
\caption{\label{F4} (Color online) Two example of temperature dependent
$(\eta/s)_\mathrm{QGP}(T)$, with which {\tt VISHNU} can simultaneously fit 
the STAR and ALICE $v_2(p_T)$ at 30-40\% centrality.}
\end{figure}
%======================================================================
%

However, Fig.~\ref{F4} shows that one can at least find two different
functional forms of $(\eta/s)_\mathrm{QGP}(T)$, with which {\tt VISHNU}
can simultaneously fit the STAR and ALICE $v_2(p_T)$ at 30-40\% centrality
bins.\footnote{$(\eta/s)(T)^{(a)}$ and $(\eta/s)(T)^{(b)}$ can also nicely
   fit the $p_T$-spectra for identified hadrons at 30-40\% centrality which,
   due to lack of data, we obtained theoretically from {\tt VISHNU}, using
   constant $\eta/s$ as input. With $p_T$ spectra and $v_2(p_T)$ fitted,
   one can also roughly fit the integrated $v_2$, since the latter is
   calculated from the former two.}
$(\eta/s)(T)^{(a)}$ monotonically increases with $T$ in the QGP phase while
$(\eta/s)(T)^{(b)}$ first increases with $T$ and then decrease with $T$ at
even higher temperature.\footnote{For the purpose of demonstration, we have
   chosen simple (even unrealistic) forms for the temperature dependence of
   $(\eta/s)(T)^{(a)}$ and $(\eta/s)(T)^{(b)}$  with 2-4 free parameters
   that can easily be fitted from spectra and $v_2$ at RHIC and LHC energies.}
Please note that the minimum values of $(\eta/s)(T)$ in case ${(a)}$ and
${(b)}$ are below 0.16, and the the maximum values are well above 
0.24.\footnote{Since below $T_\mathrm{chem}\eq165$\,MeV the fluid is 
   described microscopically, the behaviour of $(\eta/s)(T)^{(a,b)}$ shown
   in Fig.~\ref{F4} in the region $T{\,<\,}165$\,MeV is irrelevant for our
   calculation. We made no attempt to model in detail the (presently unknown)
   form of $(\eta/s)(T)$ in the phase transition region.}
Although $(\eta/s)(T)^{(b)}$ shows a smaller and subsequently negative
slope at higher temperature, it has a higher minimum value of $\eta/s$
than $(\eta/s)(T)^{(a)}$.

The fireball evolution may differ between $(\eta/s)(T)^{(a)}$ and
$(\eta/s)(T)^{(b)}$. However, both the final $v_2$ and particle spectra
are sensitive only to the time-integral of the QGP evolution, which
in both cases apparently is very similar to the evolutions at fixed values
of $\eta/s=0.16$ at RHIC energy and $\eta/s=0.22$ at LHC energy. Based on
our current analysis, utilizing only $v_2$ and particle spectra, one cannot
unambiguously determine the functional form of $(\eta/s)(T)$ and whether
the QGP fluid is more viscous or more perfect at LHC energy.

Our parametrization $(\eta/s)(T)^{(a)}$ is similar to one of the
forms used in \cite{Niemi:2011ix} whose authors studied the sensitivity
of elliptic flow measurements at RHIC and LHC to a temperature-dependent
increase of $(\eta/s)_\mathrm{QGP}$ above the quark-hadron phase transition.
The authors of \cite{Niemi:2011ix} concluded that elliptic flow measurements
at RHIC energies are insensitive to such an increase, and provided convincing
evidence that this conclusion does not depend on their treatment of the
late hadronic stage which they evolved hydrodynamically, with sudden
Cooper-Frye freeze-out, rather than microscopically as we do here. In
contrast to their work, we do find such a sensitivity; this is why in
$(\eta/s)(T)^{(a)}$ we had to lower the minimal $\eta/s$ value at low
temperatures below the constant value of 0.16 that we used in Fig.~\ref{F2},
in order to preserve the theoretical description of the $v_2(p_T)$ measured
by STAR. We have traced this difference to a sensitivity to initial
conditions: whereas Niemi {\it et al.} \cite{Niemi:2011ix} initialize
the shear viscous pressure tensor $\pi^{\mu\nu}$ at either zero or
a non-zero initial value that is independent of $\eta/s$, we use
Navier-Stokes initial conditions $\pi^{\mu\nu}=2\left(\frac{\eta}{s}(T)
\right)s\sigma^{\mu\nu}$ ($\sigma^{\mu\nu}$ is the velocity shear tensor
\cite{Song:2007fn}) where $\pi^{\mu\nu}$ increases with $T$ (i.e. towards
the fireball center) if $\eta/s$ does so. This increases the initial
transverse pressure gradients and thereby affects the final radial and
elliptic flow.

In summary, we have studied the differential and integrated $v_2$ at 
different centrality bins for 200 A GeV Au+Au collisions and 2.76 A TeV 
Pb+Pb collisions, using the hybrid model {\tt VISHNU} which describes the 
expansion of a QGP using viscous hydrodynamics and the successive evolution 
of hadronic matter with a microscopic transport model. We find that, in 
order to describe the STAR and ALICE $v_2\{4\}$ data with reaction-plane 
averaged MC-KLN initial conditions, one needs an \emph{averaged} QGP 
viscosity $\eta/s \approx 0.16$ at RHIC energies and 
$\eta/s{\,\approx\,}0.20{-}0.24$ at LHC energies. Although this result is 
in qualitative agreement with expectations from  weakly coupled QGP 
calculations \cite{Csernai:2006zz} and from recent lattice simulations
\cite{Meyer:2007ic}, both of which shows that the specific QGP shear 
viscosity increases with temperature, a more detailed analysis with 
{\tt VISHNU} utilizing a temperature dependent $(\eta/s)_\mathrm{QGP}(T)$ 
as input  shows that one cannot uniquely constrain the form of $(\eta/s)(T)$ 
by fitting the spectra and $v_2$ alone. Based on our phenomenological 
approach, it remains an open question whether the QGP fluid is more 
viscous or more perfect in the temperature regime reached by LHC 
energies.

{\it Acknowledgments:} We gratefully acknowledge fruitful discussions with
T.~Hirano, P.~Huovinen, H.~Masui, H.~Niemi, A.~Poskanzer, C.~Shen, 
R.~Snellings, and X.-N. Wang. We especially thank T.~Hirano for providing 
the new initial conditions at LHC energies, R.~Snellings for the 
experimental data shown in Figs.~\ref{F1b} and \ref{F2}, and 
H.~Buesching for the ALICE data show in Fig.~\ref{F1}a. This work was 
supported by the U.S.\ Department of Energy under Grants No. 
DE-AC02-05CH11231, DE-FG02-05ER41367, \rm{DE-SC0004286}, and (within the 
framework of the Jet Collaboration) \rm{DE-SC0004104}. We gratefully 
acknowledge extensive computing resources provided to us by the Ohio 
Supercomputer Center.

\newpage

%%%%%%%%%%%%%%%%%%%%%%%%  References %%%%%%%%%%%%%%%%%%%%%%%%%%%%%%%%%%%%%%%%%

\end{document}